\documentclass[twocolumn,showpacs,preprintnumbers,amsmath,amssymb]{revtex4}

\usepackage{color}
\usepackage{graphicx}
\usepackage{dcolumn}
\usepackage{bm}

\begin{document}

\title{Efficient qubit detection using alkali earth metal ions and a double STIRAP process}
\author{Ditte M{\o}ller}
\email{dittem@phys.au.dk}
\author{Jens L. S\o rensen}
\author{Jakob B. Thomsen}
\author{Michael Drewsen}
\affiliation{QUANTOP - Danish National Research Foundation Center for Quantum Optics,\\
Department of Physics and Astronomy, University of Aarhus, DK-8000,
Denmark.}
\date{\today }

\begin{abstract}
We present a scheme for robust and efficient projection measurement
of a qubit consisting of the two magnetic sublevels in the
electronic ground state of alkali earth metal ions. The scheme is
based on two stimulated Raman adiabatic passages (STIRAP) involving
four partially coherent laser fields. We show how the efficiency
depends on experimentally relevant parameters: Rabi frequencies,
pulse widths, laser linewidths, one- and two-photon detunings,
residual laser power, laser polarization and ion motion.
\end{abstract}

 \maketitle

\section{Introduction}
Quantum computation is a promising technique for efficient solving
of high complexity problems, which are inaccessible by classical
algorithms \cite{Nielsen00}. Currently, efforts are made within many
fields of physics in order to explore the possibility of realizing
quantum computing. Notable examples are superconducting circuits
\cite{Wallraff05,Steffen06}, semiconductors \cite{Atature06}, linear
optics with single photons \cite{Knill01,Prevedel07}, cold neutral
atoms in cavities \cite{Turchette95} and lattices
\cite{Schrader04,Mompart03,Deutsch00} as well as cold, trapped ions
\cite{Wineland03}. So far, most progress has been made in ion trap
systems
\cite{Monroe95,Schmidt-Kaler03,Brown07,Moehring07,Seidelin06,Haljan05}
where quantum gates \cite{Monroe95,Schmidt-Kaler03}, many qubit
entanglement \cite{Leibfried05,Haffner05} and quantum error
correction has been demonstrated \cite{Chiaverini04}.

\begin{figure}[b] \centering
\includegraphics[width=0.4 \textwidth]{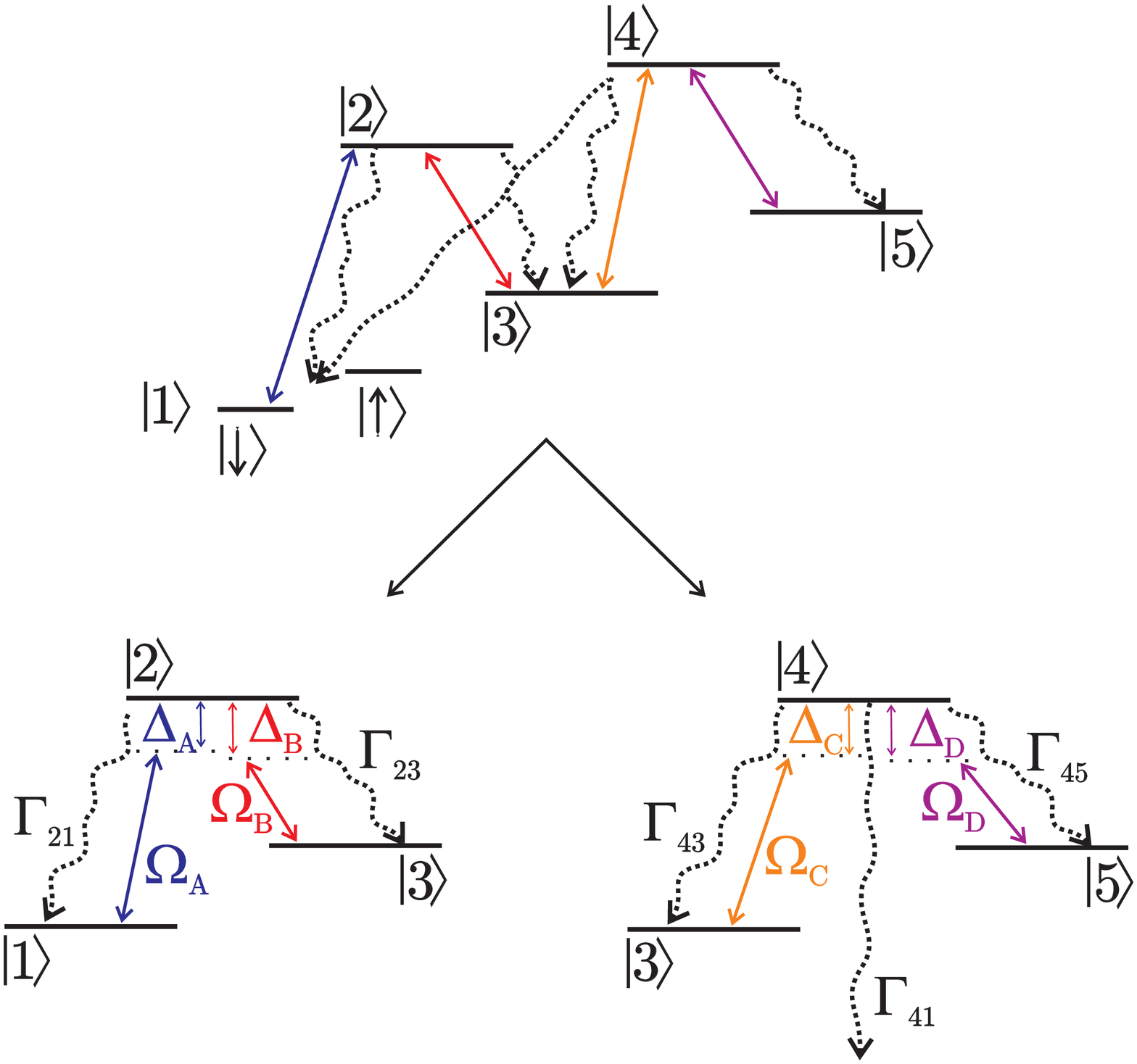}
\caption{(Color) Relevant energy levels of isotopes of Ca$^+$,
Sr$^+$, Ba$^+$ and Hg$^+$ with no nuclear spin: $\vert
1\rangle=\vert $S$_{1/2}\rangle$, $\vert
2\rangle=\vert$P$_{1/2}\rangle$, $\vert 3\rangle=\vert
$D$_{3/2}\rangle$, $\vert 4\rangle=\vert$P$_{3/2}\rangle$ and $\vert
5\rangle=\vert$D$_{5/2}\rangle$. We consider the qubit encoded in
the Zeeman sublevels, $\vert\!\!\uparrow \rangle$ and
$\vert\!\!\downarrow \rangle$ of the electronic ground state, $\vert
1\rangle$. The states $\vert 3\rangle$ and $\vert 5\rangle$ are
metastable with lifetimes varying from a fraction of a second to
many seconds depending on the ion species. The decay of these states
is neglected here. Since the scheme consists of two stages of
STIRAP, we break down the five level system into two three level
$\Lambda$ systems, each coupled by two optical fields.}
\label{fig:femniv}
\end{figure}

The qubit states most successfully implemented with trapped ions are
two hyperfine components of the $^9$Be$^+$ ground state
\cite{Monroe95} and a combination of the metastable 3D$_{5/2}$ state
and the 4S$_{1/2}$ ground state in $^{40}$Ca$^+$
\cite{Schmidt-Kaler03}. In both cases linear Paul traps are used to
confine the ions. In this work we consider the electron spin in the
alkali earth metal ion ground state as a qubit. In Fig.
\ref{fig:femniv} the electronic ground state is denoted by $\vert
1\rangle$ and the qubit basis by $\vert\!\! \uparrow\rangle$ and
$\vert\!\!\downarrow\rangle$. Single qubit operations as well as
gate operations can be performed by driving stimulated Raman
transitions between the two qubit states \cite{Moller07}, or
alternatively using controlled mechanical light forces
\cite{Staanum02}. Since solely ground states are involved, the qubit
decoherence will be limited only by ambient noise fields and ion
heating effects, rather than excited state lifetimes. Limitations
due to ambient magnetic field noise can be avoided using logical
qubits of a decoherence-free subspace
\cite{Haffner052,Kielpinski01}. In order to achieve fault tolerant
quantum computation the detection error rate must be kept low and
different error correction codes estimates allowed error rates
between $10^{-5}$ and $10^{-2}$ \cite{Knill05,Steane99}. However,
high error rates require a large overhead of qubits to encode the
error correction. In this paper we present a scheme for potentially
efficient qubit projective measurements via shelving of population
of one qubit state in the long lived metastable D$_{5/2}$ state
found in the isotopes with no nuclear spin of the alkali earth metal
ions Ca$^+$, Sr$^+$ and Ba$^+$ as well as the transition metal ion
Hg$^+$.

The qubit shelving is performed via a double Stimulated Raman
Adiabatic Passage (STIRAP) process \cite{Oreg84, Bergmann98}, as
illustrated in Fig. \ref{fig:femniv}. Initial and final states of
the shelving process are denoted $\vert\!\!\downarrow \rangle$ and
$\vert 5 \rangle$ respectively. After shelving, the atomic
population remaining in the $\vert\!\! \uparrow\rangle$ state can be
observed directly by resonantly driving the $\vert
1\rangle\rightarrow \vert 2\rangle$ and $\vert 3\rangle\rightarrow
\vert 2\rangle$ transitions and monitoring the fluorescence
\cite{Staanum04}. We use the first STIRAP process to transfer the
population from $\vert\!\!\downarrow\rangle$ to $\vert 3 \rangle$,
where we achieve the spin-state selectivity by using circularly
right handed polarized light to couple the states
$\vert\!\!\downarrow\rangle$ and $\vert 2 \rangle$. The second
process transfers the population from $\vert 3 \rangle$ to the
non-fluorescent state, $\vert 5 \rangle$. A weak externally applied
magnetic field defines the quantization axis and creates a Zeeman
splitting of $\vert\!\!\downarrow \rangle$ and $\vert\!\!
\uparrow\rangle$ (See Fig.~\ref{fig:femniv}).

STIRAP has been shown to be a robust way of adiabatically
transferring population from one quantum state to another in a three
level lambda-system using two laser pulses in a counterintuitive
order \cite{Bergmann98, Gaubatz90, Goldner94, Lawall94, Weitz94,
Cubel05, Broers92, Deiglmayr06, Sorensen06}. Compared to population
transfer via $\pi$-pulses or rapid adiabatic passage, STIRAP has the
advantage that no strict control of laser amplitude and phase is
required to maintain a high efficiency \cite{Shore92}. However,
STIRAP does require the pairs of laser pulses involved to be phase
coherent relative to each other. In practice, this means that either
the lasers must be mutually phase locked or the laser pulses
sufficiently short, such that laser decoherence is negligible. The
latter method is experimentally attractive because high obtainable
Rabi frequencies permit short pulses to be used and hence laser
phase locking is circumventable. This approach has previously been
used in the experimental realizations of STIRAP
\cite{Bergmann98,Sorensen06}. Also note that the two STIRAP stages
can be decoupled within the lifetime of intermediate metastable
state, $\vert 3 \rangle$. Of course decoherence is always at play.
Hence, in the numerical calculations below we assume finite laser
linewidths and only partially coherent lasers.

The goal of this work is to theoretically investigate the shelving
process and identify the parameters of importance to its efficiency.
In Sec.~\ref{sec:general} we first make some general remarks on the
STIRAP-process. This analytical treatment is valid for any three
level $\Lambda$ system and it serves to develop the criteria for
maintaining adiabaticity using Gaussian shaped pulses. We then turn
to numerical simulations of the density matrix evolution for the
full five level system. Here, we consider the roles of one- and
two-photon detunings of the Raman resonances, laser linewidths,
laser pulse widths, Rabi frequencies, stray laser light, laser
polarizations and ion motion. The results are presented in Sec.'s
\ref{sec:simulationsias} and \ref{sec:polarizationrequirements}.
Here we use as an example the $^{40}$Ca$^{+}$ ion, but, apart from
different decay rates and wavelengths, our treatment is also valid
for Sr$^+$, Ba$^+$ and Hg$^+$ ions with no nuclear spin. In
Sec.~\ref{sec:fullsim} we present simulations taking all effects
into account and finally we conclude in Sec. \ref{sec:conclusion}.

\section{STIRAP with Gaussian pulses \label{sec:general}}
\begin{figure}[b]
\includegraphics[width=0.45\textwidth]{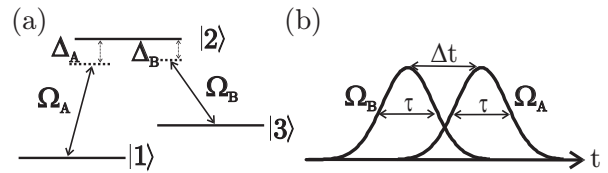}
\caption{\label{fig:pumpstoke}(a) Three level lambda-system with a
pump laser (A) and a Stokes laser (B) applied. The two laser fields
have Rabi frequencies $\Omega_A$ and $\Omega_B$ and they are detuned
$\Delta_A$ and $\Delta_B$ from resonance. (b) STIRAP pulse sequence.
$\Delta t$ is the pulse separation and $\tau$ is the full width at
$1/e$ height for both pulses.}
\end{figure}

Let us first consider one STIRAP process involving the three atomic
basis states $(\vert 1\rangle, \vert 2\rangle, \vert 3\rangle)$. In
this section we will neglect spontaneous decay of the $\vert
2\rangle$ state, since this is of no importance to the adiabaticity underlying STIRAP. Two
mutually coherent and monochromatic laser fields $A$ (pump field)
and $B$ (Stokes field) couples $\vert 1\rangle$ and $\vert 3\rangle$
to $\vert 2\rangle$ (See Fig.~\ref{fig:pumpstoke}(a)). The
interaction Hamiltonian for this system in the rotating wave
approximation is given by

\begin{equation}
H_I (t)=\frac{\hbar}{2} \left (
\begin{array}{ccc}
0 & \Omega_A (t) & 0 \\
\Omega_A (t) & 2\Delta_{A} & \Omega_B (t) \\
0 & \Omega_B (t) & 2(\Delta_{A}-\Delta_{B})
\end{array}
\right ) \label{hamil}
\end{equation}
where $\Delta_{A}$ and $\Delta_{B}$ are the detunings of the two
applied fields having the real, time dependent Rabi frequencies,
$\Omega_A (t)$ and $\Omega_B (t)$ respectively. These parameters are
defined as in \cite{Bergmann98}. On two-photon resonance,
$\Delta_{A}-\Delta_{B}=0$, diagonalizing (\ref{hamil}) yields the
eigenvalues
\begin{equation}
\omega^{\pm}=\frac{1}{2}(\Delta_A\pm\sqrt{\Delta_{A}^{2}+\Omega_{A}^{2}+\Omega_{B}^{2}})
\hspace{0,2cm}, \hspace{1cm}\omega^{d}=0,\label{Eq. eigenvalues}
\end{equation}
corresponding to eigenstates
\begin{eqnarray}
 \vert +\rangle & = & \sin (\alpha)\sin(\beta)|1\rangle + \cos(\beta)|2\rangle -\cos(\alpha)\sin(\beta)|3\rangle , \notag \\
\vert -\rangle & = & \sin (\alpha)\cos(\beta)|1\rangle -\sin(\beta)|2\rangle +\cos(\alpha)\cos(\beta)|3\rangle ,\notag \\
\vert d\rangle & = & \cos (\alpha)|1\rangle - \sin(\alpha)|3\rangle,
\label{Eq. eigenstates}
\end{eqnarray}
with $\tan(\alpha)=\frac{\Omega_{A}(t)}{\Omega_{B}(t)}$ and
$\tan(\beta)=$\mbox{$\sqrt{-\frac{\omega^{-}}{\omega^{+}}}$}. The
$|d\rangle$-state has zero interaction energy and hence it is
decoupled from the light. As a result adiabatic following of this
state does not populate of the potentially short-lived $\vert
2\rangle$-state. Now, with all population initially in $\vert
1\rangle$ and only the Stokes field applied, this initial state is
$|d\rangle$. Adiabatically decreasing $\Omega_B$ while increasing
$\Omega_A$ corresponds to a variation of $\tan(\alpha)$ from $0$ to
$\infty$. The pulse sequence thus changes the dressed state
$|d\rangle$, from $\vert 1\rangle$ to $\vert 3 \rangle$.

Since STIRAP requires adiabatic following of atoms while remaining
in the $|d\rangle$-state, we will now consider the criterion for
this to happen. We assume essentially Fourier limited optical pulses
with a Gaussian time dependence. These give rise to the time varying
Rabi frequencies
\begin{eqnarray}
\Omega_A (t) & = & \Omega_{A,0} \exp\left [-\left (\frac{t-\Delta t/2}{\tau /2} \right )^2 \right], \notag\\
\Omega_B (t) & = & \Omega_{B,0} \exp\left [-\left (\frac{t+\Delta t/2}{\tau
/2} \right )^2 \right]\label{gausspulses}
\end{eqnarray}
where the peak Rabi frequencies are $\Omega_{A,0}$ and
$\Omega_{B,0}$, $\Delta t$ is the pulse separation and $\tau$ is the
full width at $1/e$ height, assumed to be the same for both pulses
(See pulse sequence in Fig.~\ref{fig:pumpstoke}(b)). We parameterize
the problem in terms of the scaled time, $\theta=\sqrt{8}t/\tau$,
scaled pulse separation,
 $\eta=\sqrt{2}\Delta t/\tau$ and Rabi frequency asymmetry $r=\Omega_{A,0}/\Omega_{B,0}$. Hence
\begin{eqnarray}
\Omega_A (\theta) & = & \Omega_{A,0} \exp\left [-1/2\left ( \theta -\eta \right )^2 \right], \notag \\
\Omega_B (\theta) & = & \Omega_{B,0} \exp\left [-1/2\left (\theta+\eta \right
)^2 \right], \label{scalegausspulses}
\end{eqnarray}
and
\begin{equation}
\vert d \rangle =-\frac{ r^{-1/2} e^{-\eta\theta} \vert 1\rangle -
r^{1/2} e^{\eta\theta} \vert 3\rangle }{\sqrt{r
e^{2\eta\theta}+r^{-1} e^{-2\eta\theta}}}. \label{dstate}
\end{equation}
Adiabaticity requires the rate of change of the wave function to be
small compared to the energy separation between the dressed state
eigenvalues,
\begin{equation}
\left\vert \frac{d}{ dt} \vert d\rangle \right\vert \ll \vert \omega_{\pm} \vert.
\label{adiabaticity}
\end{equation}
The rate of change of the wave function (\ref{dstate}) is
parameterized by the Bloch sphere polar angle in the $\lbrace \vert
1\rangle, \vert 3\rangle \rbrace$ basis, which from (\ref{Eq.
eigenstates}) and (\ref{dstate}) is found to be
$\alpha=\arctan[r\exp(2\eta\theta)]$. The wave function rate of
change is then
\begin{equation}
\frac{d\alpha}{d\theta}=\frac{2\eta}{r e^{2\eta\theta} + r^{-1}
e^{-2\eta\theta}}. \label{thetadot}
\end{equation}
This should be compared to the dimensionless eigenvalue of the
energetically closest dressed state, $\vert-\rangle$. On two-photon
resonance and in the limit of large single photon detuning,
$\Omega_{A,0},\Omega_{B,0}\ll\vert\Delta_{A}\vert$, the eigenvalue
is given by
\begin{equation}
\lambda_- = \frac{\omega^{-}\tau}{\sqrt{8}}= - 2\Lambda \left (r
e^{2\eta\theta} + r^{-1} e^{-2\eta\theta} \right ) \exp\left[-\left
(\theta^2 + \eta^2 \right )\right ], \label{brightenergy}
\end{equation}
We can now formulate an adiabaticity criterion, which should be
fulfilled at all times during the pulse sequence, where the Rabi frequencies are appreciable. The criterion states that
\begin{equation}
A(\theta,\eta,r)\equiv\frac{\vert\eta\vert \exp \left[\theta^2 + \eta^2\right
]}{\left (r e^{2\eta\theta} + r^{-1} e^{-2\eta\theta} \right )^2}
\ll \Lambda, \label{adcrit}
\end{equation}
where we identify the parameter relevant for maintaining adiabaticity
\begin{equation}
\Lambda=\frac{\Omega_{A,0} \Omega_{B,0} }{16\sqrt{2}\vert\Delta_{A}\vert}\tau. \label{adparam}
\end{equation}
On inspection $\Lambda$ is simply proportional to the product of the
maximum achievable Raman Rabi frequency and the duration of the
pulse sequence. Hence, Eq. (\ref{adcrit}) simply states that many
Raman Rabi cycles should take place during the time span of the
STIRAP process, analogous to the result derived in
\cite{Bergmann98}.

\begin{figure}
\includegraphics[width=0.45\textwidth]{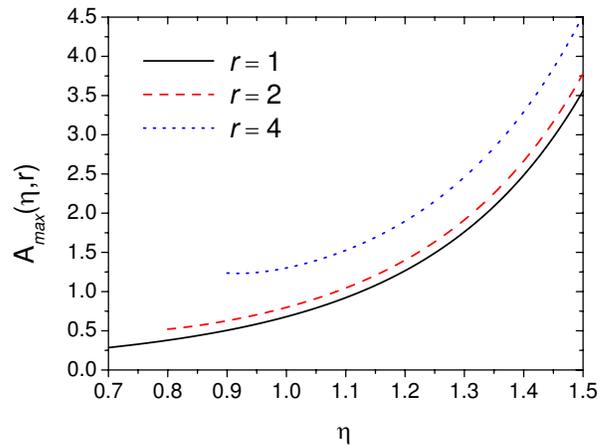}
\caption{\label{amaxfig}(Color) Maximum value of the adiabaticity
function, $A_{max}(\eta,r)=A(\theta_{max},\eta,r)$ as function of
STIRAP pulse separation, $\eta$, for $r=1$, $r=2$ and $r=4$. For $r>1$ a solution does not exist for smaller $\eta$ values, hence the truncation of the corresponding curves. The maximum is found by solving Eq. (\ref{eq:thetaopt}).}
\end{figure}
The timing of the STIRAP sequence is contained in the function
$A(\theta,\eta,r)$. For a range of parameters this has a maximum as
function of time, $A_{max}(\eta,r)=A(\theta_{max},\eta,r)$, where
$\theta_{max}$ solves the equation
\begin{equation}
2\tanh(2\eta\theta_{max}+\ln r)=\theta_{max}/\eta.
\label{eq:thetaopt}
\end{equation}
For an efficient STIRAP process to occur, given a certain value of
$r$, we obviously want to choose the pulse separation, $\eta$, so
that $A_{max}$ is small and obeys the inequality (\ref{adcrit}).

In Fig. \ref{amaxfig} we plot $A_{max}$ as function of $\eta$ for
values, where Eq. (\ref{eq:thetaopt}) has a solution. The three
curves correspond to $r=\lbrace 1,2,4\rbrace$ and it should be noted
that $A_{max}(\eta,r)=A_{max}(\eta,r^{-1})$. From the graphs we
infer that values of $\Lambda$ in excess of 3-5 should be obtained
in order to ensure efficient population transfer. Adiabaticity is
favored when the pulse separation is decreasing toward $\eta =0.7$
and when the Rabi frequencies are balanced, corresponding to $r$
approaching unity. In an experimental situation \cite{Sorensen06}
residual light illuminating the ions before and after the STIRAP
pulse sequence can become a severe limitation to the population
transfer efficiency. This will be discussed in more detail in
Sec.~\ref{sec:backgroundlight}, where one conclusion is that a short
duration of the experiment is needed in order to avoid population
repumping due to residual light driving a parasitic Raman resonance.
Hence, we introduce the time duration of the experiment, $\Theta$,
which when inserted in Eq.~(\ref{dstate}) provide the population
transfer efficiency:
\begin{equation} P_3(\Theta)=\left [1+r^{-2}
\exp(-4\eta\Theta) \right ]^{-1}. \label{transferprob}
\end{equation}
Obviously, (\ref{transferprob}) shows that the transfer efficiency
grows with increasing $\eta$, due to the tails of the Gaussian
pulses. However, for larger $\eta$ fulfilling the adiabaticity
criteria (\ref{adcrit}) becomes difficult and we hence expect to
find an optimum value of the pulse separation, $\eta_{opt}$, which
depends only on $r$ and $\Lambda$. With respect to $r$,
$P_3(\Theta)$ grows for high values, while adiabaticity requires
values close to unity and therefore one might expect an optimal
value of $r$ above $1$ with the exact value depending on $\Lambda$. To
optimize the transfer efficiency with respect to $\eta$ and $r$, we
solve the optical Bloch equations for the three level $\Lambda$
system. Details of our derivation of the Bloch equations can be
found in the next section.

In order to make our Bloch equation solutions general, we again
ignore the decay from the short lived state $\vert 2\rangle$ and we
assume perfectly coherent laser fields. The Bloch equations are
integrated numerically for one-photon detunings of 4 GHz. These
detunings are chosen sufficiently large, so that for Rabi
frequencies in the 100 MHz range, the results are independent of the
specific values of the detunings. It is checked numerically that
increasing the detunings even more has no impact on the solutions.
In this limit our results can be considered general and system
independent.

\begin{figure} [t]
\includegraphics[width=0.45\textwidth]{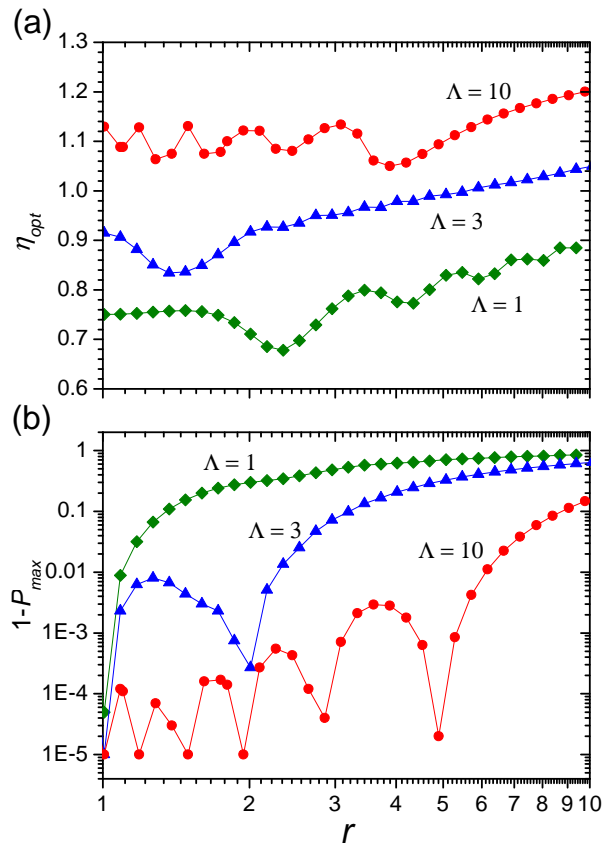}
\caption{\label{etamaxfig} (Color) (a) Optimum pulse separation,
$\eta_{opt}$ as function of Rabi frequency ratio,
$r=\Omega_{A,0}/\Omega_{B,0}$. As indicated, the three curves
correspond to different values of $\Lambda$, given by $\Lambda=1$
(\textcolor[rgb]{0,0.6,0}{{\textbf{$\blacklozenge$}}}), $\Lambda=3$
(\textcolor[rgb]{0,0,1}{{\textbf{$\blacktriangle$}}}) and
$\Lambda=10$ (\textcolor[rgb]{1,0,0}{{\textbf{$\bullet$}}}).
One-photon detunings are $\Delta_{A}/2 \pi=\Delta_{B}/2 \pi= 4$ GHz
and all results are obtained with $\tau=2$ $\mu$s. (b) The transfer
inefficiency $1-P_{max}$ corresponding to the values of $\eta_{opt}$
calculated in (a). The oscillations of the curves are attributed to
Rabi dynamics and it should be noted that the curves are symmetric
with respect to letting $r \rightarrow r^{-1}$.}
\end{figure}

For fixed values of $r$ and $\Lambda$ the population transfer is
computed as a function of $\eta$, and the optimum value,
$\eta_{opt}$, corresponding to the maximum population transfer,
$P_{max}$, is found. Examples of such curves can be found in Sec.
\ref{sec:effectofpulsedelay}, where we analyze the sensitivity of
the population transfer with respect to fluctuations of $\eta$. Fig.
\ref{etamaxfig}(a) shows $\eta_{opt}$ as function of $r$ for
$\Lambda = \lbrace 1,3,10 \rbrace$. From the curves it is found that
for values of $r$ far from unity, a larger pulse separation is
preferable. As mentioned above, this is attributed to the wings of
the larger amplitude Gaussian pulse, which must decay $(r<1)$ -or
grow $(r>1)$ sufficiently before the pulse sequence is terminated.
This is also the reason why $\eta_{opt}$ grows from about 0.7 for
$\Lambda = 1$ to about 1.1 for $\Lambda = 10$. Fig.
\ref{etamaxfig}(b) shows the population transfer inefficiency,
$1-P_{max}$, associated with the values of $\eta_{opt}$ computed in
Fig. \ref{etamaxfig}(a). From these curves it is obvious that
balanced Rabi frequencies are crucial for an efficient population
transfer when the adiabaticity criterion (\ref{adcrit}) is only
marginally fulfilled, as it is the case for $\Lambda=1$. As
$\Lambda$ grows, we see that higher asymmetries in the Rabi
frequencies can be tolerated. This is in accordance with the
discussion of Fig. \ref{amaxfig}. It should be noted that the
inclusion of finite laser linewidths tends to make $\eta_{opt}$
smaller as a result of decoherence mechanisms explained in Sec.
\ref{sec:laserlinewidth}. For the remaining simulations in this
paper we have sought to use optimum values of $\eta$ according to
the reasoning above.

Although Fig. \ref{etamaxfig}(b) shows that the fidelity of qubit
detection in general can be high for $\Lambda=10$, it should be
noted that the difference between a detection error of
$10^{-3}$ and $10^{-5}$ becomes very important when error correction
protocols are considered \cite{Nielsen00}. Hence, optimizing the
fidelity with respect to $r$, according to Fig. \ref{etamaxfig}(b),
could be crucial if the qubit detection scheme should be utilized
for scalable quantum computation.

\section{Optical Bloch equations}\label{sec:blochequations}
The previous section developed analytical criteria for maintaining
adiabaticity when Gaussian pulses are used and when the one-photon
detunings are much larger than Rabi frequencies. We now turn to
numerical simulations of the detection scheme taking into account
the full dynamics and using realistic experimental parameters. We
solve the optical Bloch equations describing the dynamics of the
five levels depicted in Fig.~\ref{fig:femniv}, but ignore magnetic
sublevels. We stay within the dipole approximation and define the
Rabi frequencies\begin{eqnarray}
\Omega_{A}=\mu_{12}E_A/\hbar \\
\Omega_{B}=\mu_{32}E_B/\hbar \\
\Omega_{C}=\mu_{34}E_C/\hbar \\
\Omega_{D}=\mu_{54}E_D/\hbar
\end{eqnarray} where $\mu_{ij}$ are the dipole matrix elements and $E_i$ laser
field amplitudes. The laser fields with frequencies $\omega_{ij}$
are not necessarily on resonance and we introduce detunings
$\Delta_{ij}=\frac{\mathcal{E}_{i}-\mathcal{E}_{j}}{{\hbar}}-\omega_{ij}$
with respect to the atomic energy levels $\{\mathcal{E}_{i}\}$. With
the notation of Fig.~\ref{fig:femniv} we define
$\Delta_A=\Delta_{21}, \Delta_B=\Delta_{23}, \Delta_C=\Delta_{43}$
and $\Delta_D=\Delta_{45}$. Furthermore, we apply the rotating wave
approximation to arrive at the interaction Hamiltonian for the
system
\begin{small}
\begin{equation}
H=\sum_{i=1}^{5} \hbar \omega_i \rho_{ii}+(\hbar \Omega_{A}
\rho_{12} + \hbar \Omega_{B} \rho_{32} + \hbar \Omega_{C} \rho_{34}
+ \hbar \Omega_{D} \rho_{54} + h.c.),
\end{equation}
\end{small}where $\omega_i=\mathcal{E}_i/\hbar$ and the density operator
elements are $\rho_{ij} = \vert j\rangle\langle i\vert$. Spontaneous
emission is introduced as decay terms in the density matrix
elements. We neglect spontaneous decay from the metastable
$|3\rangle$ and $|5\rangle$ states because their lifetimes are
significantly longer than the simulation time (See lifetimes of the
different ion species in appendix). Phase fluctuations of the lasers
are introduced as decay of the coherences and ion micromotion as a
harmonic modulation of the detunings.

In the simulations we use the specific wavelengths ($\lambda_A=397$
nm, $\lambda_B=866$ nm, $\lambda_C=850$ nm, $\lambda_D=854$ nm) and
spontaneous decay rates ($\Gamma_{21} /2\pi=21$ MHz, $\Gamma_{23}
/2\pi=1.7$ MHz, $\Gamma_{41} /2\pi=22$ MHz, $\Gamma_{43} /2\pi=0.18$
MHz, $\Gamma_{45} /2\pi=1.6$ MHz) for the $^{40}$Ca$^+$ ion, but
this apart the calculations are also valid for isotopes of Sr$^+$,
Ba$^+$ and Hg$^+$ with no nuclear spin (See relevant parameters for
the most abundant isotopes in appendix). We use Gaussian pulses as
defined in Eq.~(\ref{gausspulses}) with pulse widths $\tau_{i}=2\mu
s$. All other parameters are varied depending on which
investigations are made. When there is no argument for a different
choice we use Rabi frequencies $\Omega_{i,0}/2\pi=100$ MHz and
detunings $\Delta_i/2\pi=600$ MHz. The delay between pulses is
chosen optimal, which in the simulations presented is either $\Delta
t = 1.2\mu s$ or $\Delta t = 1.3\mu s$. The theory of
Sec.~\ref{sec:general} applies when $|\Omega_{i,0}|\ll|\Delta_{i}|$,
which is the case in the majority of the simulations. In these cases
we specify the values of $\eta$ and $\Lambda$ to make comparison
with Sec.~\ref{sec:general} straightforward. For the parameters
mentioned above $\eta=0.85$ or $0.92$ and $\Lambda=9.3$. All
parameters for the simulations are mentioned in the figure captions
and hence omitted in the text.

To recapitulate, the detection scheme consists of two STIRAP
processes. The first transfers population from $|1\rangle$ to
$|3\rangle$ applying the $A$ and $B$-fields as pump and Stokes
field, respectively. The second process between $|3\rangle$ and
$|5\rangle$ uses $C$ as the pump field and $D$ as the Stokes field.

We first investigate the effect of pulse delay
(Sec.~\ref{sec:effectofpulsedelay}), laser detunings
(Sec.~\ref{Sec:laserdetunings}), laser linewidth
(Sec.~\ref{sec:laserlinewidth}), pulse width
(Sec.~\ref{sec:roleofpulsewidths}) and residual light
(Sec.~\ref{sec:backgroundlight}). All these simulations are
performed on the latter transition because this transition is
subject to spontaneous decay out of the sub-system and thus has a
higher sensitivity to non-adiabaticity. An experimental
investigation of this transition can be found in \cite{Sorensen06}.

The first STIRAP stage is more sensitive to ion motion because the
two laser fields used have very different wavelengths and the
Doppler shift induced by the ion motion therefore gives a large
two-photon detuning. Since this stage is responsible for the
internal state selection, it is also important to investigate the
effect of laser polarization errors and the resulting depletion of
the wrong qubit state. The simulations of ion micromotion (Sec.
\ref{sec:ionmotion}) and polarization errors (Sec.
\ref{sec:polarizationrequirements}) are therefore performed on the
first STIRAP transition. The simulations are made with all the
population initially in $|1\rangle$ or $|3\rangle$ and we find the
transfer efficiency as the final population in $|3\rangle$ or
$|5\rangle$ for the first or the second STIRAP stage respectively.
We assume that the calculated transfer efficiency is equivalent to
the qubit detection efficiency, and we do thereby not focus on
signal to noise issues of the fluorescence detection of the
non-shelved state $|\!\!\uparrow\rangle$.

\section{Simulations ignoring atomic spin}\label{sec:simulationsias}
In this chapter we assume perfect laser polarization and hence
ignore irrelevant magnetic sublevels of the ions. As mentioned
above, we start out by considering the latter
$|3\rangle\rightarrow|4\rangle\rightarrow|5\rangle$ STIRAP process.

\subsection{Effect of pulse delay}\label{sec:effectofpulsedelay}
\begin{figure}[htbp]
\centering
\includegraphics[width=0.45 \textwidth]{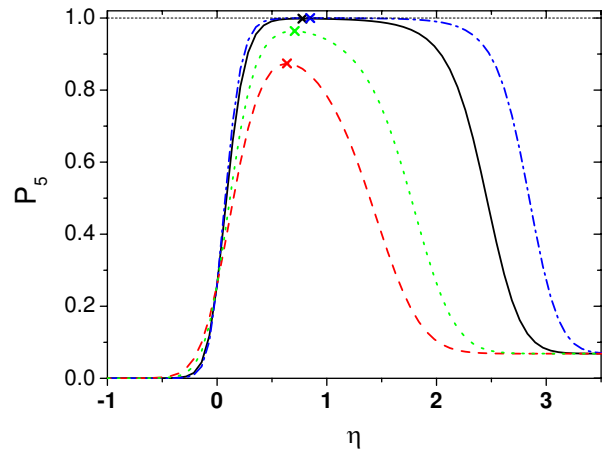}
\caption{(Color) Transfer efficiency as a function of delay between
pulses for different choices of peak Rabi frequencies:
(\mbox{\textcolor[rgb]{1,0,0}{\textbf{- - -}}})
 $\Omega_{C,0}=\Omega_{D,0}=2\pi\times 10$ MHz,
 (\mbox{\textcolor[rgb]{0,1,0}{\large{\textbf{$\cdot$$\cdot$$\cdot$}}}})
 $\Omega_{C,0}=\Omega_{D,0}=2\pi\times 20$ MHz,
 (\mbox{\textbf{\large{---}}}) $\Omega_{C,0}=\Omega_{D,0}=2\pi\times 100$ MHz,
 (\mbox{\textcolor[rgb]{0,0,1}{\large{\textbf{-{$\cdot$}-}}}})
 $\Omega_{C,0}=\Omega_{D,0}=2\pi\times 300$ MHz. Positive
delay correspond to the counter intuitive pulse sequence, where the
laser pulse of field $D$ arrives before the laser pulse of field $C$.
Parameters: $\tau_C=\tau_D=2$ $\mu$s and
$\Delta_C=\Delta_D=0$.\label{fig:Delayscan}}
\end{figure}

The first parameter we investigate is the delay between the two
pulses and  in Fig. \ref{fig:Delayscan} we show the transfer
efficiency as a function of $\eta$ for various peak Rabi frequencies
$\Omega_{C,0}=\Omega_{D,0}$. Both laser fields are on resonance.
The simulations show that for Rabi frequencies above
$\Omega_{C,0}=\Omega_{D,0}=2\pi\times 100$ MHz the transfer
efficiency is close to unity and insensitive to fluctuations in the
delay ($P_{_{5}}>0.999$ for fluctuations of $\eta$ less than 0.2
around the optimal value of 0.85). For smaller Rabi frequencies the
transfer efficiency decreases and the sensitivity to delay
fluctuations increases. The plateau of $P_5 \simeq 0.07$ for
positive $\eta$ values is the result of optical pumping by the pump
field, $\Omega_C$. For $\eta
>0$ this pulse is applied last, and the Stokes field cannot repump
population.

We wish to investigate how changes in the delay between pulses
influence the transfer efficiency when it is not diminished by other
effects. The results of Fig.~\ref{fig:Delayscan} are therefore
calculated for zero one-photon detuning. In this limit the theory of
Sec. \ref{sec:general} is not valid, but in the rest of the work
one-photon detunings considerably larger than Rabi frequencies will
be used and hence the analytical theory can be used to interpret the
numerical simulations.

The results presented in Fig.~ \ref{fig:Delayscan} still show the
same trends as the analytical theory. High Rabi frequencies are
needed to maintain adiabaticity and, as indicated by the crosses,
the delay giving maximum transfer efficiency is found to grow with
increasing Rabi frequencies, consistent with Fig. \ref{etamaxfig}.

\subsection{Laser detunings} \label{Sec:laserdetunings}
\begin{figure}[t] \centering
\includegraphics[width=0.5 \textwidth]{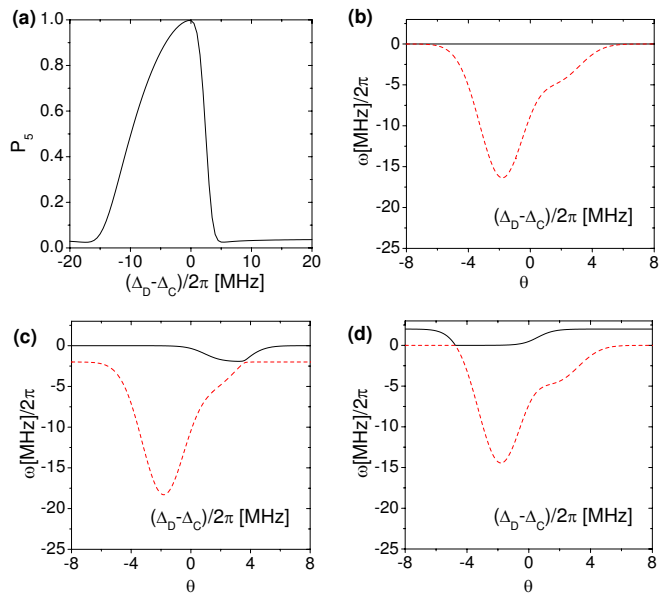}
\caption{(Color)(a) Transfer efficiency as a function of two-photon
detuning. Evolution of eigenvalues as a function of time is shown in
(b):$\Delta_D-\Delta_C=0$ MHz, (c)
$\Delta_D-\Delta_C=2\pi\times 2$ MHz and (d)
$\Delta_D-\Delta_C=-2\pi\times 2$ MHz . In (b), (c) and (d) we use
the signatures: (\mbox{\textbf{\large{---}}}) $\omega^D$,
(\mbox{\textcolor[rgb]{1,0,0}{\large{\textbf{- - -}}}}) $\omega^-$.
Parameters used for all graphs: $\Omega_{C,0}=2\pi\times 100$ MHz,
$\Omega_{D,0}=2\pi\times 200$ MHz, $\Delta_{C}=2\pi\times 600$ MHz,
$\tau_{C}=\tau_{D}=2$ $\mu$s and $\Delta t =1.2$ $\mu$s. ($\Lambda=18.5$, $\eta=0.85$).}
\label{fig:egenv}
\end{figure}

To reduce the probability of incoherent diabatic excitations we
introduce large one-photon detunings of the lasers. Thus, we now
investigate the sensitivity of STIRAP to one- and two-photon
detunings.
\begin{figure}[htbp]
\centering
\includegraphics[width=0.5 \textwidth]{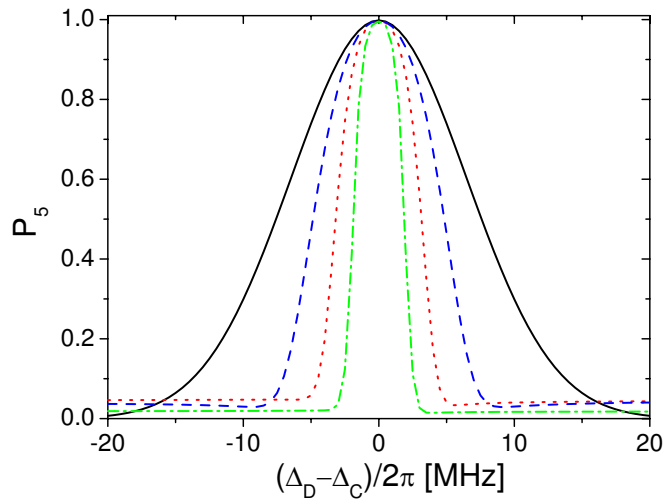}
\caption{(Color) Transfer efficiency as a function of two-photon
detuning ($\Delta_D-\Delta_C$) of the C and D lasers for different
choices of one-photon detuning, $\Delta_C$. The simulations are made
with $\Omega_{C,0}=\Omega_{D,0}=2\pi\times 100$ MHz,
$\tau_{C}=\tau_{D}=2$
 $\mu$s and $\Delta t =1.2$ $\mu$s $(\eta=0.85)$. (\mbox{\textbf{\large{---}}})
$\Delta_C= 0$ MHz, (\mbox{\textcolor[rgb]{0,0,1}{\textbf{- -
-}}}) $\Delta_C=2\pi\times 300$ MHz $(\Lambda = 18.5)$,
 (\mbox{\textcolor[rgb]{1,0,0}{\large{\textbf{$\cdot$$\cdot$$\cdot$$\cdot$}}}})
 $\Delta_C=2\pi\times 600$ MHz $(\Lambda = 9.3)$,
 (\mbox{\textcolor[rgb]{0,1,0}{\large{\textbf{-{$\cdot$}-{$\cdot$}}}}})
 $\Delta_C=2\pi\times 1200$ MHz $(\Lambda = 4.6)$.}
\label{fig:Frekvensscan}
\end{figure}
The simulations, presented in Fig.~\ref{fig:Frekvensscan}, show that
increasing the one-photon detuning does not limit the transfer
efficiency as long as we are close to the two-photon resonance
$\Delta_D-\Delta_C=0$. This criterion gets stricter as we increase
the one-photon detuning. If we require $P_5> 0.99$ the demand on
two-photon detuning is $\Delta_D-\Delta_C\pm$ $2\pi\times 0.5$ MHz
when $\Delta_C= 2\pi\times 1200$ MHz as estimated from the
dash-dotted curve of Fig. \ref{fig:Frekvensscan}. For
$\Delta_C=2\pi\times 600$ MHz, maintaining a two-photon resonance
within $2\pi\times 1$ MHz is sufficient as seen from the dotted
curve of Fig. \ref{fig:Frekvensscan}. An efficient transfer can thus
be maintained in spite of a drift from two-photon resonance making
STIRAP robust with respect to small frequency drifts of the involved
lasers. A closer look at Fig.~\ref{fig:Frekvensscan} reveals that
the spectra are not exactly symmetric with respect to the sign of
the two-photon detuning. With unbalanced Rabi frequencies,
$\Omega_{D,0}=2\Omega_{C,0}$ (See. Fig.~\ref{fig:egenv}(a)), the
asymmetry becomes more evident. For small negative two-photon
detunings the transfer efficiency is much higher than for the
corresponding positive two-photon detuning. This effect is due to
diabatic transfer between $|d\rangle$ and the energetically closest
bright state $|-\rangle$. As discussed in Sec. \ref{sec:general}
this is likely when $|d\rangle$ and $|-\rangle$ are nearly
degenerate.
The eigenvalues found in Eq. (\ref{Eq. eigenvalues}) are valid only
on two-photon resonance, and the solutions are more complicated in
the situation, where this condition is not met. A general expression
for the eigenvalues has been derived in \cite{Fewell97} and is
plotted in Fig.~\ref{fig:egenv}(b-d). Here, we show the eigenvalues
of $|d\rangle$ (solid curves) and $|-\rangle$ (dashed curves) as a
function of time for three different choices of two-photon detuning.
In the case where $\Delta_D-\Delta_C=0$ MHz (See
Fig.~\ref{fig:egenv}(b)) the eigenvalues of course coincide when the
Rabi frequencies are zero before and after the pulses, but in this
case no diabatic transfer will occur as no coupling is present. For
positive and negative detunings ($\Delta_D-\Delta_C=2\pi\times 2$
MHz, Fig.~\ref{fig:egenv}(c) and $\Delta_D-\Delta_C=-2\pi\times 2$
MHz, Fig.~\ref{fig:egenv}(d)) we see avoided crossings leading to a
probability for diabatic transfer to the $|-\rangle$-state. Such a
transfer leads to population of the $|4\rangle$-state, which decays
rapidly, and the population mainly ends up in $|1\rangle$. When the
energy splitting between the $|d\rangle$ and the $|-\rangle$ state
is small, maintaining adiabaticity requires high coupling strengths.
For a positive two-photon detuning the avoided crossing occurs late
in the sequence, as shown in Fig.~\ref{fig:egenv}(c). Here the
coupling strength is relatively small due to the asymmetry
$\Omega_{D,0}=2\Omega_{C,0}$, hence the probability of diabatic
transfer is large. With a negative two-photon detuning the avoided
crossing occurs early, as shown in Fig.~\ref{fig:egenv}(c). The
coupling is stronger at this time and the probability of diabatic
transfer is thus smaller. This difference gives rise to the
asymmetry of the two-photon spectrum in Fig.~\ref{fig:egenv}(a).

\subsection{Laser linewidths}\label{sec:laserlinewidth}
\begin{figure}[htbp]
\centering
\includegraphics[width=0.5 \textwidth]{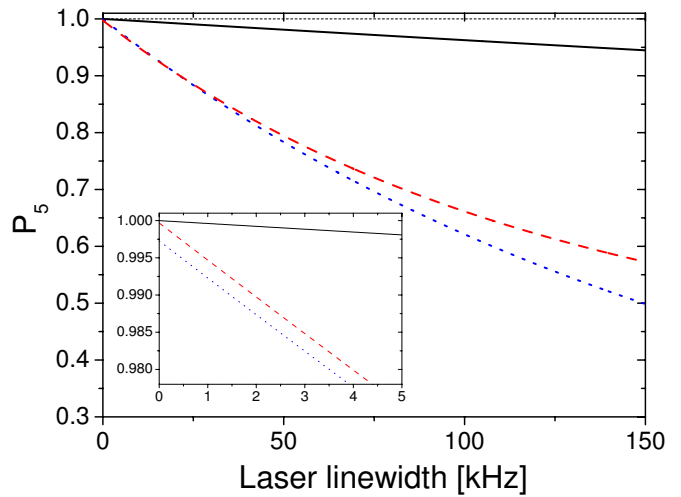}
\caption{(Color) Transfer efficiency as a function of the linewidth
of the C and D laser fields for different choices of Rabi
frequencies. The simulations use $\tau_{C}=\tau_{D}=2$ $\mu$s,
$\Delta t =1.2$ $\mu$s ($\eta=0.85$) and
$\Delta_{C}=\Delta_{D}=2\pi\times 600$ MHz.
(\mbox{\textbf{\large{---}}})
$P_{5}=1/3+2/3\exp({-3\gamma_{35}\tau^{2}/16\Delta t})$,
(\mbox{\textcolor[rgb]{1,0,0}{\large{\textbf{- -
-}}}})$\Omega_{C,0}=\Omega_{D,0}=2\pi\times 300$ MHz
($\Lambda=83.3$),
 (\mbox{\textcolor[rgb]{0,0,1}{\large{\textbf{$\cdot$$\cdot$$\cdot$}}}})
$\Omega_{C,0}=\Omega_{D,0}=2\pi\times 100$ MHz ($\Lambda=9.3$).} \label{fig:liniebredde}
\end{figure}

Laser phase fluctuations lead to dephasing between the three states
involved and  may therefore affect STIRAP. This dephasing effect has
been studied thoroughly in \cite{Ivanov04}, where the STIRAP
transfer efficiency has been shown to depend only on the dephasing
rate between $|3\rangle$ and $|5\rangle$, $\gamma_{35}$.
\begin{equation}
P_{5}=\frac{1}{3}+\frac{2}{3}e^{-3\gamma_{35}\tau^{2}/16\Delta
t},\label{dephasing}
\end{equation}
for Gaussian pulses within the adiabatic limit, not taking decay
from $|4\rangle$ into account. In Fig.~\ref{fig:liniebredde} we
present simulations showing the exact influence of the laser phase
fluctuations parameterized by the laser linewidth. Also shown as the
full curve is Eq.~(\ref{dephasing}).
The curves show that the transfer efficiency is strongly limited by
the laser phase fluctuations and because of the non-zero decay from
$|4\rangle$ the situation is even worse than predicted by
Eq.~(\ref{dephasing}). For high Rabi frequencies the coupling is
stronger and therefore the limitations due to dephasing are less
pronounced. To maintain a transfer efficiency above $0.99$ we find
laser linewidths required to be below $2\pi\times 1.5$ kHz for
$\Omega_{C,0}=\Omega_{D,0}=2\pi\times 100$ MHz and below $2\pi\times
2$ kHz for $\Omega_{C,0}=\Omega_{D,0}=2\pi\times 300$ MHz. It should
be noted that by laser linewidth we mean the frequency fluctuations
averaged over the time duration of the experiment. This is not
necessarily the steady state laser linewidth. For pulse sequences of
roughly 10 $\mu$s duration, as are employed in our simulations,
laser linewidths of a few kHz are not unrealistic for most laser
systems.

\subsection{Role of pulse widths}\label{sec:roleofpulsewidths}
\begin{figure}[htbp]
\centering
\includegraphics[width=0.5 \textwidth]{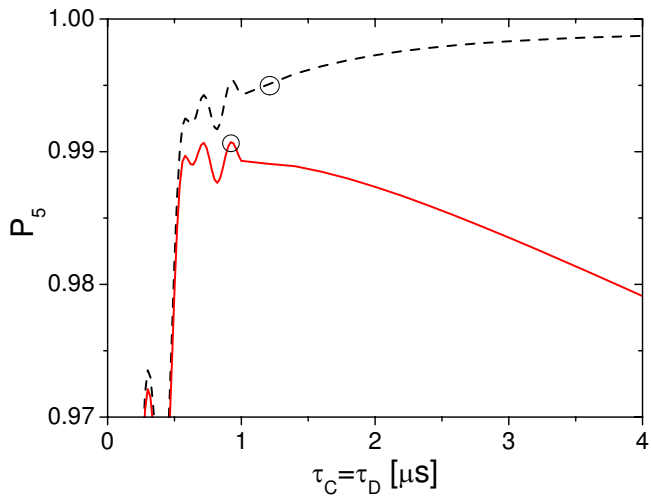}
\caption{(Color) Transfer efficiency as a function of pulse width of
the $C$ and $D$ laser fields for $2\pi\times 2$ kHz
(\textcolor[rgb]{1,0,0}{\large{\textbf{---}}}) as well as no laser
linewidth (\textbf{\large{- - -}}) The delay between the pulses is
given by $\eta =0.85$, laser detunings are
$\Delta_{C}=\Delta_{D}=2\pi\times 600$ MHz and peak Rabi frequencies
$\Omega_{C,0}=\Omega_{D,0}=2\pi\times 100$ MHz ($\Lambda=9.3$).}
\label{fig:pulse}
\end{figure}

As discussed above, the laser phase fluctuations limit the transfer
efficiency, making it preferable to use short pulses, since this
typically reduces the effective laser linewidth. Short pulses,
however, require higher Rabi frequencies to maintain the
adiabaticity. In Fig.~\ref{fig:pulse}, we show the transfer
efficiency as a function of pulse width in the case of no laser
linewidth as well as the case of a $2\pi\times 2$ kHz linewidth. As
expected, for the case of no linewidth we find an increasing
efficiency as we increase the pulse widths because the criterion for
adiabaticity is better fulfilled for larger pulse widths as found in
(\ref{adparam}). We indicate the point where we exceed $0.995$ on
the graph. When we introduce a $2\pi\times 2$ kHz laser linewidth an
optimum pulse width is found as indicated on the graph.

\begin{figure}[htbp]
\centering
\includegraphics[width=0.5 \textwidth]{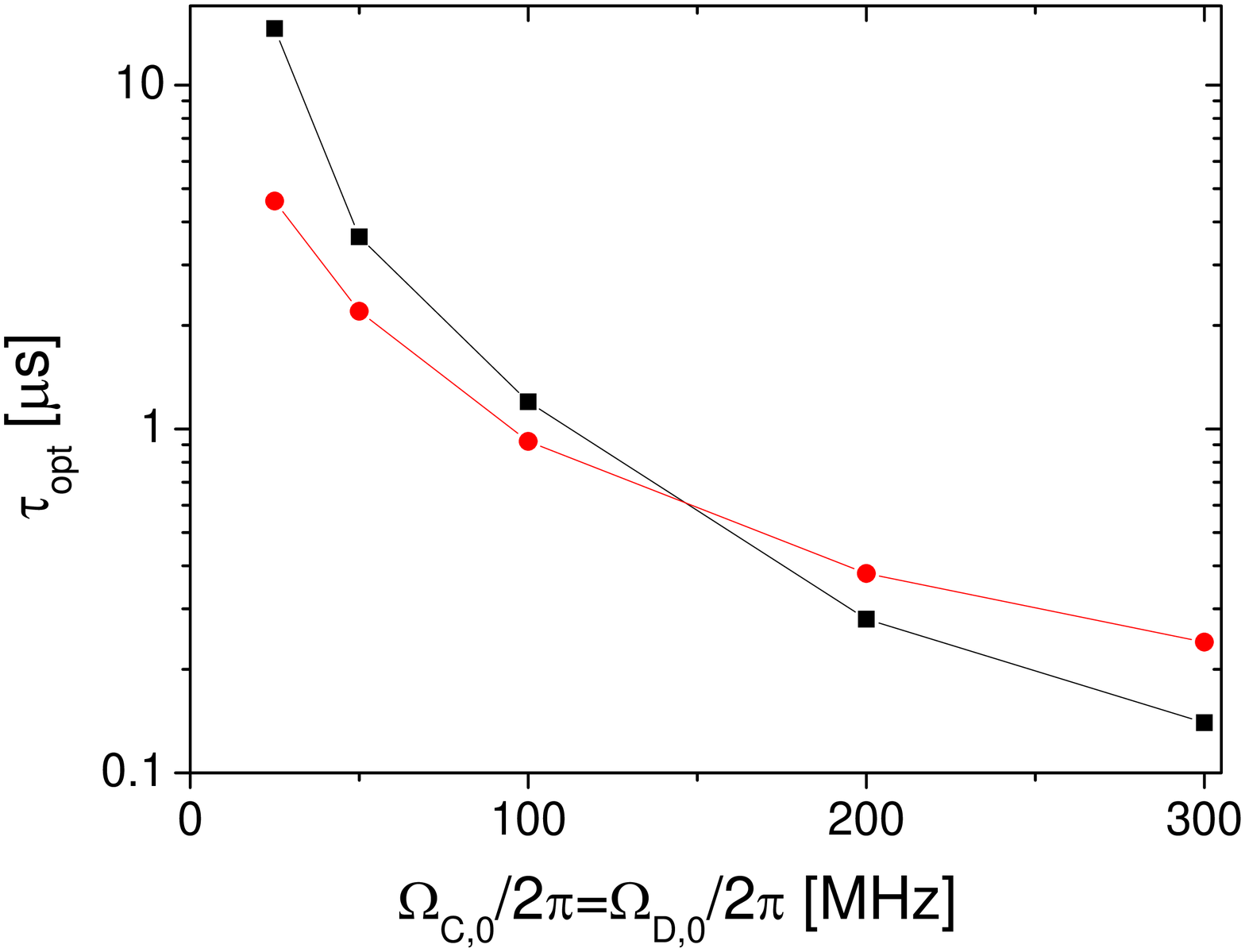}
\caption{(Color) Optimal pulse width as a function of the Rabi
frequencies of the $C$ and $D$ laser fields for $2\pi\times 2$ kHz
(\textcolor[rgb]{1,0,0}{{\textbf{$\bullet$}}})  as well as no laser
linewidth ({\tiny{\textbf{{$\blacksquare$}}}}). The delay between
the pulses is given by $\eta =0.85$, laser detunings are
$\Delta_{C}=\Delta_{D}=2\pi\times 600$ MHz.} \label{fig:pulseopt}
\end{figure}

This optimum depends on the Rabi frequencies of the lasers and the
dependency is pictured in Fig.~\ref{fig:pulseopt}. In the case of no
linewidth, instead of the optimum, we plot the pulse widths required
to exceed a $0.995$ transfer efficiency. This result is compared to
the optimum pulse width, when lasers have $2\pi\times2$ kHz
linewidth. Both curves show that short pulses are preferred as Rabi
frequencies grow. However, the mutual dephasing of the lasers
involved, results in smaller optimum pulse widths when finite laser
linewidths are introduced. For higher Rabi frequencies, the two
curves cross, so that shorter pulse widths apparently are preferred
for the perfectly coherent lasers. This is a result of the
artificial limit $0.995$ chosen to define the black curve. We have
assumed the laser linewidths to be constant over the range of pulse
durations involved. More realistically one would expect the laser
linewidths to grow proportional to $(\tau_{C,D})^{\alpha}$, where
the power, $\alpha$, is in the range $0.5 \rightarrow 1$ for a phase
diffusion process.

\subsection{Residual light}\label{sec:backgroundlight}
In the first generation of STIRAP experiments beams of atoms passed
through two stationary laser beams \cite{Gaubatz90,Bergmann98}.
Here, the detection region was spatially well separated from the
STIRAP region and in this case no residual light was present. When
we consider the situation with stationary ions and pulsed laser
beams, the situation is less favorable. In the case of a constant
background level of laser light, the ions do not start out in an
exact dark state and hence population transfer may be limited by the
resulting finite population of the short lived states $\vert
2\rangle$ and $\vert 4\rangle$. Moreover, since the residual light
is able to excite the Raman resonance between the states $\vert
3\rangle$ and $\vert 5\rangle$, repumping of the transferred
population can take place after ended STIRAP pulse sequence. With a
small background level of light this Raman resonance will be narrow
and hence the limitation of population transfer is most severe near
two-photon resonance as experimentally found in \cite{Sorensen06}.
Assuming Rabi frequencies of $2\pi\times100$ MHz and extinction
ratio $\approx10^{-4}$ of the optical pulse generators, we arrive at
residual Rabi frequencies in the MHz range, which can easily repump
a significant portion of the shelved population on two-photon
resonance. In this section we address this problem and establish
upper limits on the peak Rabi frequencies involved.

\begin{figure}[htbp] \centering
\includegraphics[width=0.5 \textwidth]{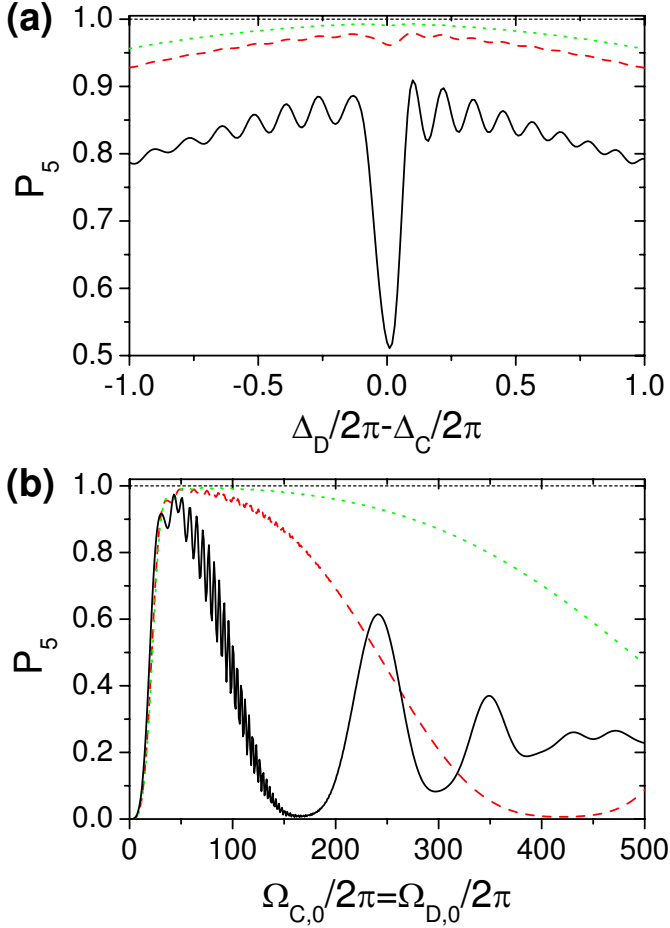}
\caption{(Color) (a) Transfer efficiency as a function of two-photon
detuning with $\Omega_{C,0}=\Omega_{D,0}=2\pi\times 100$ MHz and
$\Delta_{C}=2\pi\times 600$ MHz. $\Lambda=9.3$. (b) Transfer
efficiency as the peak Rabi frequencies are varied, simulated with
$\Delta_{C}=\Delta_{D}=2\pi\times 600$ MHz. The different curves
represent different fractions of residual light:
(\mbox{\textcolor[rgb]{0,1,0}{\large{\textbf{$\cdot$$\cdot$$\cdot$}}}})
$\Omega_{\textrm{off},i}/\Omega_{i,0}=0.01$,
(\mbox{\textcolor[rgb]{1,0,0}{\large{\textbf{- - -}}}})
$\Omega_{\textrm{off},i}/\Omega_{i,0}=0.02$ and
(\mbox{\large{\textbf{---}}})
$\Omega_{\textrm{off},i}/\Omega_{i,0}=0.05$. Parameters used in
simulations: $\tau_C=\tau_D=2$ $\mu$s, $\Delta t=1.2$ $\mu$s,
yielding $\eta=0.85$. The total time duration is $20$ $\mu$s and
laser linewidth has been ignored.} \label{fig:offpower}
\end{figure}

We denote the Rabi frequency of the residual light
$\Omega_{\textrm{off},i}$, $i=\{A,B,C,D\}$. We first look at the
transfer efficiency as a function of the two-photon detuning. When
no residual light is present the transfer efficiency is optimal on
two-photon resonance as shown previously in
Fig.~\ref{fig:Frekvensscan}, but this is no longer the case when we
introduce a fraction of residual light,
$\Omega_{\textrm{off},i}/\Omega_{i,0}$. In
Fig.~\ref{fig:offpower}(a) we present the transfer efficiency as a
function of two-photon detuning for different fractions of residual
light. As we increase this fraction we see an increasing loss of
population on two-photon resonance. With
$\Omega_{\textrm{off},i}/\Omega_{i,0}$ less than 0.02 (dashed
curves) this loss is only a few percent relative to the maximum
transfer efficiency. However, already with
$\Omega_{\textrm{off},i}/\Omega_{i,0}=0.05$ (solid curve) the loss
has grown to nearly 50 \% near two-photon resonance, rendering the
detection scheme useless in this case.

As mentioned earlier, residual light is typically present as a
fraction of the peak Rabi frequencies, and hence very high Rabi
frequencies are not necessarily preferable. In
Fig.~\ref{fig:offpower}(b) we show the transfer efficiency as a
function of the peak Rabi frequencies in the STIRAP pulses for
different values of the fraction of residual light. Increasing the
Rabi frequencies increases the transfer efficiency until the
residual light limits the transfer. The maximum transfer efficiency
is found to be $0.99$ around the Rabi frequencies,
$\Omega_{C,0}=\Omega_{D,0}=2\pi\times 50$ MHz with a residual light
level less than $\Omega_{\textrm{off},i}/\Omega_{i,0}=0.02$. For
even higher Rabi frequencies we observe Rabi-floppings between
$|3\rangle$ and $|5\rangle$.

Experimentally $\Omega_{i,0}/2\pi=100$ MHz can be achieved focusing
a beam with a mW power level to a $\mu$m spot size. Switching of the
lasers with acusto-optical modulators or electro-optical modulators
can be done on ns timescale reducing the power level to a few
hundred nW which correspond to
$\Omega_{\textrm{off},i}/\Omega_{i,0}=0.01$. This is represented by
the green dotted curve in Fig.~\ref{fig:offpower}(b), which
obviously reveals the limitations set by residual light. The
extinction ratio $10^{-4}$ correspond to the situation where the
lasers are extinguished by one optical pulse generator. Applying
additional AOM's and EOM's in succession will reduce the residual
light level, and as a result, the limitation due to this light can be smaller as well, at the expense of increased experimental complexity.

\subsection{Ion motion}\label{sec:ionmotion}
The ions have so far been considered as having no motion with
respect to the laser fields. We consider the ions to be confined in
a linear Paul trap \cite{Ghosh95}, where they experience an
effective confining potential due to RF- and DC-fields. The ion
motion can be derived to consist of a three-dimensional harmonic
motion with frequencies $\omega_x$, $\omega_y$ and $\omega_z$ called
the secular motion \cite{Ghosh95}. In addition the RF-field drives
an oscillation with the RF-frequency, $\Omega_{RF}$. This
oscillation is called the ion micromotion. The secular frequencies
are typical much smaller than $\Omega_{RF}$ and hence STIRAP
transfer efficiency is mainly limited by the micromotion. The effect
of micromotion is simulated by modulating the detunings as
\begin{equation}
\Delta_i(t)=\Delta_i(0)-v\frac{2\pi}{\lambda_i}\cos(\Omega_{RF} t).
\end{equation}

\begin{figure}[t]\centering
\includegraphics[width=0.5 \textwidth]{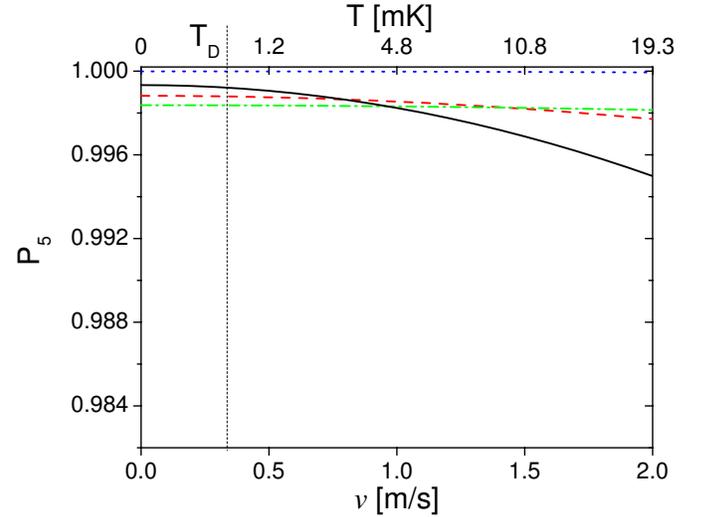}
\caption{(Color) Transfer efficiency as a function of the
micromotion velocity. The different curves show results for
different detunings:
(\textcolor[rgb]{0,0,1}{\large{\textbf{$\cdot$$\cdot$$\cdot$}}})
 $\Delta_A=\Delta_B=$ 0 MHz,
 (\textbf{\large{---}}) $\Delta_A=\Delta_B= 2\pi\times 300$ MHz $(\Lambda=18.5)$,
(\textcolor[rgb]{1,0,0}{\large{\textbf{- - -}}})
 $\Delta_A=\Delta_B=2\pi\times 600$ MHz $(\Lambda=9.3)$ and
(\textcolor[rgb]{0,1,0}{\large{\textbf{-{$\cdot$}-}}})
 $\Delta_A=\Delta_B=2\pi\times 1200$ MHz $(\Lambda=4.6)$. Parameters used for
 simulations: $\tau_A=\tau_B=2$ $\mu$s, $\Delta t=1.3$ $\mu$s $(\eta=0.92)$,
 $\Omega_{A}=\Omega_{B}=2\pi\times 100$ MHz and
$\Omega_{RF}=2\pi\times 16.8$ MHz. The temperature of the ion is
defined as T=$m v^2/k_{B}$, where $m$ is the ion mass and $k_B$
Boltzmann's constant. The Doppler temperature for Ca$^+$, T$_D$, is
indicated by the vertical line.}
  \label{fig:mikro}
\end{figure}

We only consider micromotion in one dimension with maximum velocity
$v$. This classical treatment of micromotion is justified by quantum
Monte Carlo simulations where the ion motion has been quantized
showing no visible difference between classical and quantized motion
\cite{Thomsen05}. Another argument for this classical simplification
is the strength of the laser fields making $\Omega_{i,0} \gg
\Omega_{RF}$. As discussed previously, the STIRAP transfer
efficiency is more sensitive to two-photon than one-photon
detunings.  If we assume co-propagating laser fields and take the
$^{40}$Ca$^+$ ion as an example, the two-photon detuning will be
strongly modulated in the case of the $\vert 1\rangle$-$\vert
2\rangle$-$\vert 3\rangle$ transition where the two laser fields
have a large wavelength difference $(\Delta\lambda/\lambda)_{123} =
0.54$, while the $\vert 3\rangle$-$\vert 4\rangle$-$\vert 5\rangle$
transition experiences a much smaller modulation of the two-photon
detuning $(\Delta\lambda/\lambda)_{345} = 0.005$. This means that
the first transition, in the case of $^{40}$Ca$^+$, is strongly
limited by micromotion while the second remains virtually
unaffected.

In Fig.~\ref{fig:mikro} we show the transfer efficiency as a
function of the micromotion velocity for different choices of
detuning. The simulations show that when we increase the one-photon
detuning the micromotion velocity becomes an increasingly limiting
factor. For small micromotion velocities we see that the transfer
efficiency is higher for small one-photon detunings. This is due to
a decreased sensitivity to a non-vanishing two-photon detuning as
discussed in Sec. \ref{Sec:laserdetunings}. For higher velocities
the one-photon detuning is strongly modulated. This results in
sidebands, of which the high frequency component comes close to
resonance, hereby inducing real transitions to the short lived
$\vert 2\rangle$-state. This effect gets more evident when the
one-photon detuning is close to $v\frac{2\pi}{\lambda_i}$, which for
our parameters varies between $0$ and $100$ MHz.
Fig.~\ref{fig:mikro} shows small sensitivity to micromotion for very
small and very large values of $\Delta_A=\Delta_B$, while the
$\Delta_A=\Delta_B=2\pi\times$300 MHz trace shows a rapid drop as a
function of $v$. For the chosen parameters a micromotion velocity
below $1$ m/s ensures a transfer efficiency above $0.998$. From the
kinetic energy of the ion a temperature can be defined through the
relation $\frac{1}{2}k_B$T$=\frac{1}{2}mv^2$. Hence, in the case of
$^{40}$Ca$^+$ a velocity below $1$ m/s correspond to a temperature
below 4.8 mK. When the ion is Doppler laser cooled, temperatures
below the so-called Doppler temperature, T$_D$, cannot be obtained.
For $^{40}$Ca$^+$ T$_D=0.5$ mK as indicated by the vertical line on
Fig. \ref{fig:mikro}. This is well below the required 4.8 mK and
Doppler cooling is thus sufficient to achieve a high transfer
efficiency. Doppler temperatures for other relevant ions can be
found in Table \ref{tab:iondata}.

\section{Polarization requirements}\label{sec:polarizationrequirements}
\begin{figure}[t] \centering
\includegraphics[width=0.45 \textwidth]{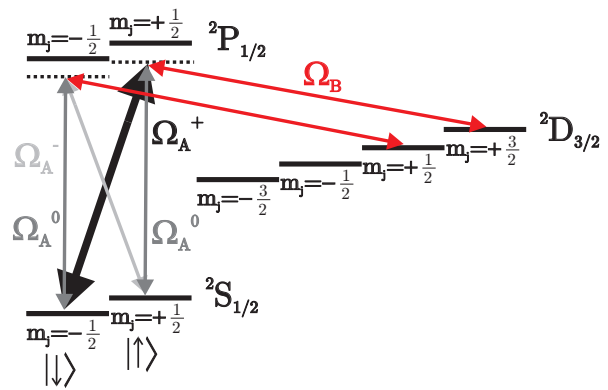}
\caption{(Color) Zeeman sublevels of the three states involved in
the first STIRAP process. The $S_{1/2}$ and $P_{1/2}$ states
are coupled by a laser field with a $\sigma_+$ polarized component
($\Omega_A ^+$), a $\sigma_-$ polarized component ($\Omega_A ^-$) and a
$\pi$ polarized component ($\Omega_A ^0$). The $D_{3/2}$ and
$P_{1/2}$ states are coupled by a laser field ($\Omega_B$) with a
pure $\sigma_-$ polarization.} \label{fig:zeemanniv}
\end{figure}
\begin{figure}[htbp]
\centering
\includegraphics[width=0.45\textwidth]{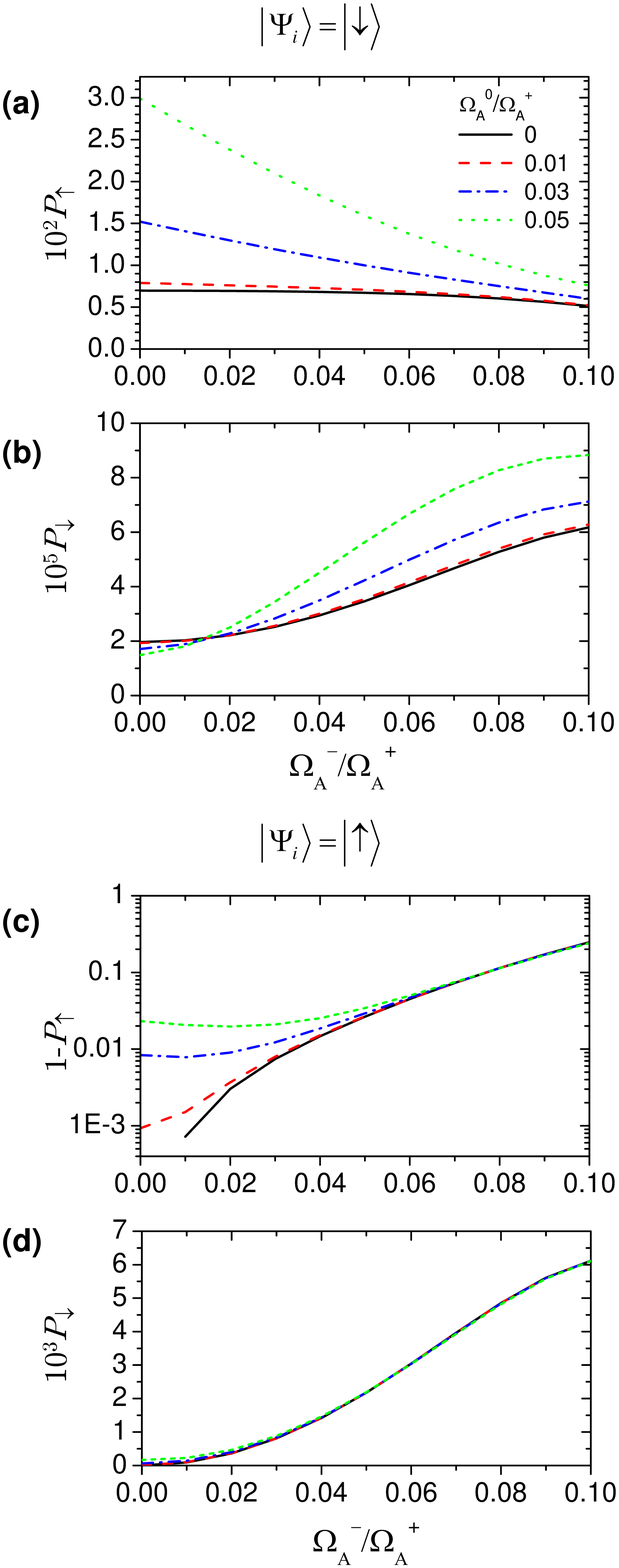}
\caption{(Color) Effect of polarization error and error in magnetic
field direction. Populations of the qubit states
$\vert\!\!\downarrow \rangle$, $P_-$, and $\vert\!\!\uparrow
\rangle$, $P_+$, plotted as a function of the relative $\sigma_-$
polarization component, $\Omega_A ^-/\Omega_A ^+$. The various
curves on each plot correspond to different magnetic field direction
errors, parameterized by the relative $\pi$ polarization component,
$\Omega_A ^0/\Omega_A ^+$. (\textbf{\large{---}}) $\Omega_A
^0/\Omega_A ^+=0$, (\textcolor[rgb]{1,0,0}{\large{\textbf{- - -}}})
$\Omega_A ^0/\Omega_A ^+=0.01$,
(\textcolor[rgb]{0,0,1}{\large{\textbf{-{$\cdot$}-{$\cdot$}}}})
$\Omega_A ^0/\Omega_A ^+=0.03$ and
(\textcolor[rgb]{0,1,0}{\large{\textbf{$\cdot$$\cdot$$\cdot$}}})
$\Omega_A ^0/\Omega_A ^+=0.05$. The upper two graphs, (a) and (b),
correspond to all initial population in the $\vert\!\!\downarrow
\rangle$ qubit state, while the lower two graphs, (c) and (d),
correspond to all initial population in $\vert\!\!\uparrow \rangle$.
The parameters for the simulations are $\Omega_{A,0} ^+
=\Omega_{B,0}=2\pi\times 300$ MHz, $\Delta_A=\Delta_B=2\pi\times
300$ MHz, $\tau=2$ $\mu$s and $\Delta t=1.3$ $\mu$s. ($\Lambda=167$,
$\eta=0.92$).} \label{fig:polerr}
\end{figure}
Until this point Zeeman sublevels have been ignored. However, in
order to investigate the effect of laser polarization, deviating
from purely circular, we now introduce all of the Zeeman sublevels
of the first stage STIRAP. The optical Bloch equations for the full
eight level system are derived analogous to the equations in
Sec.~\ref{sec:blochequations}. We consider the field $A$ to have
both $\sigma_{\pm}$ polarization components as well as a possible
$\pi$ component in case of magnetic field errors. The second STIRAP
field, $B$, is assumed to have perfect $\sigma_-$ polarization. We
denote the $\sigma_+$ polarized component of $\Omega_A$ by $\Omega_A
^+$, the $\sigma_-$ polarized component by $\Omega_A ^-$ and the
$\pi$ polarized component by $\Omega_A ^0$ (See
Fig.~\ref{fig:zeemanniv}).

Efficient qubit projection requires all population moved from
$\vert\!\!\downarrow \rangle$ to the $D_{3/2}$ levels while leaving
the $\vert\!\!\uparrow \rangle$ population unchanged. In order to
test these criteria we simulate the problem with two different
initial states corresponding to all population in
$\vert\!\!\downarrow \rangle$ and $\vert\!\!\uparrow \rangle$
respectively. The results are shown in Fig.~\ref{fig:polerr}. First
we focus on the situation, where all initial population is in the
$\vert\!\!\downarrow \rangle$ qubit state. Ideally the final
populations in both qubit states should be vanishing for our scheme
to work. The results, shown in Fig.~\ref{fig:polerr} (a) and (b),
reveal that detection errors can be kept below 0.01 provided the
magnetic field can be controlled accurately enough to keep $\Omega_A
^0/\Omega_A ^+$ below 0.02. However, when all initial population is
in the $\vert\!\!\uparrow \rangle$ state, the situation is worse.
From Fig.~\ref{fig:polerr} (c), we find that the population in the
$\vert\!\!\uparrow \rangle$ state is pumped out as the relative
amplitude of the $\sigma_-$ polarized component is increasing. This
is of course undesirable. Hence, in order to keep qubit detection
errors below 0.01 for an arbitrary initial state, we must demand
that $\Omega_A ^0/\Omega_A ^+ < 0.02$ and $\Omega_A ^-/\Omega_A ^+ <
0.04$ with the parameters used in this simulation.

\section{Full simulation} \label{sec:fullsim}
\begin{figure}[t]
\centering
\includegraphics[width=0.45\textwidth]{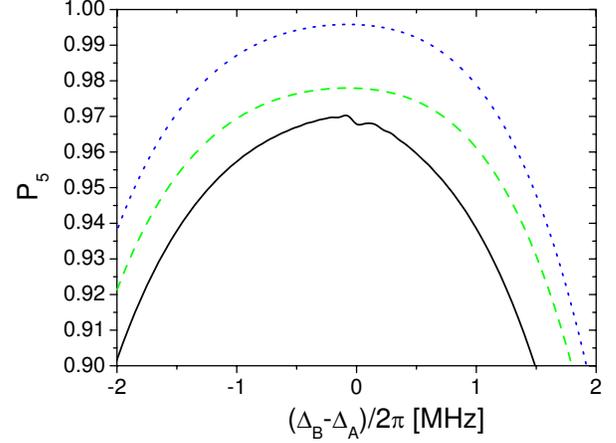}
\caption{(Color) Full simulation of the two stage STIRAP. The plot
shows the population transfer as a function of two-photon detuning
of the first STIRAP stage. Parameters are:
$\Delta_A=\Delta_C=\Delta_D=2\pi\times 600$ MHz,
$\Omega_A=\Omega_B=\Omega_C=\Omega_D=2\pi\times 100$ MHz,
$\tau_A=\tau_B=\tau_C=\tau_D=2 \mu$s. All laser linewidths are
$2\pi\times 2$ kHz and the trap is characterized by
$\Omega_{RF}=2\pi\times 16.7$ MHz and $v=0.4$ m/s, corresponding to
a temperature of 0.8 mK. The STIRAP's have $\Delta t=1.2 \mu$s
corresponding to $\eta=0.85$ and $\Lambda=9.3$ for both stages. The
two STIRAP stages are separated by 10 $\mu$s and the overall
simulation time is 30 $\mu$s. The Rabi frequencies associated with
the residual power are 0.01 relative to the peak Rabi frequencies
for all four fields. The curves correspond to:
(\textbf{\large{---}}) all effects on,
(\textcolor[rgb]{0,1,0}{\large{\textbf{- - -}}}) residual power set
to zero but laser linewidth on, and
(\textcolor[rgb]{0,0,1}{\large{\textbf{{$\cdot$}{$\cdot$}{$\cdot$}}}})
both laser linewidth and residual power set to zero.}
\label{fig:fdobb}
\end{figure}

Finally, we present the results of a simulation of the full two
stage STIRAP population transfer including all of the above
described effects except polarization errors. The parameters
relevant to the simulation can be found in the caption of Fig.
\ref{fig:fdobb}. The figure shows the population transfer efficiency
as a function of the two-photon detuning, $\Delta_B-\Delta_A$, for
the first STIRAP stage. Assuming that the ions can be kept cooled
close to the Doppler temperature, we have chosen a temperature of
0.8 mK due to the micromotion  for this simulation. The resulting
peak velocity of 0.4 m/s results in a loss of population transfer of
less than 0.01 according to the discussion of Sec.
\ref{sec:ionmotion}. The two-photon detuning of the second STIRAP
stage has been chosen to be zero and the residual power level is
chosen to yield a Rabi frequency of 0.01 relative to the peak Rabi
frequencies for all four optical fields.

From Fig. \ref{fig:fdobb} we find, when all effects are taken into
account, a maximum transfer efficiency of 0.97 for $\Delta_B
-\Delta_A=-2\pi\times 0.09$ MHz. This corresponds to the full, black
curve. In order to identify the processes limiting this efficiency
we show with the green, dashed curve the resulting efficiency when
the residual optical power is set to zero. Comparison with the full
curve shows that a loss in efficiency of almost 0.01 can be
attributed to repumping by residual laser light. The blue, dotted
curve of Fig. \ref{fig:fdobb} shows the transfer efficiency when
both residual light and laser linewidths are set to zero. Here a
maximum transfer efficiency of 0.996 is found near two-photon
resonance indicating that the finite laser linewidth is indeed a
deleterious effect to STIRAP. The linewidth of 2 kHz assigned to all
lasers in the simulation results in a loss of transfer efficiency of
nearly 0.02. In order to reduce the effect of laser phase
fluctuations shorter pulse durations and consequently higher Rabi
frequencies must be used.

The simulation illustrates the importance of efficient switching on
and off laser power in the STIRAP process. Unless the lasers
involved in the two STIRAP stages are pairwise phase locked one
obviously needs fairly high intensities of the optical pulses and
hence the requirements to the residual laser power become more
stringent. However, as can be seen from Fig. \ref{fig:Delayscan}, it
is noted that the scheme is insensitive to \textit{fluctuations} in
the laser power as long as the transitions involved are well
saturated and adiabaticity is maintained. Experimentally, high Rabi
frequencies can be achieved at the wavelength listed in the
appendix, and the complication of phase locking can thus be avoided
\cite{Sorensen06}.

\section{Summary and Conclusions}\label{sec:conclusion}
In summary, we analyze the scheme where two magnetic sublevels of
the S$_{1/2}$ ground state of alkali earth metal ions are considered
as qubit states. We have shown that qubit projection can be
effectively performed by electronically shelving the population of
one magnetic sublevel via a double STIRAP process. Hereby the
population is transferred to the metastable D$_{5/2}$ state and
remaining population of the other qubit state can be detected by
driving the strong Doppler cooling transition of the ions and
spatially monitoring the fluorescence. In order to selectively
shelve only one qubit state we apply appropriately polarized light
in the first STIRAP process.

We have established the formalism describing the system consisting
of the 5 energetically lowest levels of the alkali earth metal ions
and we model the STIRAP processes using Gaussian shaped pulses from
partially phase coherent lasers. Specific simulations have been
performed by solving the optical Bloch equations for the
$^{40}$Ca$^+$ ion. From these simulations we conclude that our
scheme is indeed feasible with a projection efficiency exceeding 99
{\%}, provided laser linewidths can be kept below 1.5 kHz over the
timescale of the pulse sequence. The short timespan required to
avoid decoherence of the lasers makes it necessary to have high Rabi
frequencies in the $100's$ of MHz regime in order to maintain
adiabaticity. However, since we are driving strong, dipole allowed
transitions, such high Rabi frequencies can be achieved by using
moderately focused lasers with 5-10 mW power.

In order to avoid unwanted excitations, we assume all lasers to be
detuned roughly 0.5 - 1 GHz from the one-photon resonance. In the
limit of this detuning dominating over Rabi frequencies, we have
identified the adiabaticity criteria for the ions when Gaussian
light pulses are applied. Given an available laser power and hence
certain Rabi frequencies, we find an optimum pulse separation
depending only on the parameters, $\Lambda$ and $r$. Within
variations of 20 \% we find the optimum pulse separation to be $\tau
/\sqrt{2}$, where $\tau$ is the $1/e$ full width of the Gaussian
pulses.

In contrast to early STIRAP experiments, where atomic beams and
stationary laser beams were used, our system involves stationary
atoms and laser pulses. Since the detection zone and the STIRAP zone
are spatially identical for stationary ions, we have to take into
account the finite extinction ratio of the optical pulse generators
involved in the experiment. Due to a finite extinction ratio, a
fraction of the peak Rabi frequencies of the STIRAP pulses will be
present before and after the pulse sequence. As a result, shelved
population can be repumped, and consequently the shelving
efficiency reduced. Clearly mechanical shutters will efficiently
block laser beams, exposing the ions, but unfortunately such
shutters need several tens of microseconds to activate. On the
shorter time scale AOM's and EOM's can be employed with extinction
ratios around $1:10^{4}$. Hence, we have analyzed the role of
residual light at the level of $10^{-2}$ in Rabi frequency, and this
was found to restrict the shelving efficiency severely if the Rabi
frequency exceeded $2\pi\times 100$ MHz. In order for the detection
scheme to be more efficient, efforts must be made to extinguish the
light even further. This could be achieved by introducing more
AOM's or EOM's in succession.

Motion in the ion trap is included in the analysis, and found to be
unimportant close to the Doppler cooling limit. For calcium ions
this is 0.5 mK.

Finally, we have analyzed the effect of impure polarization of
light, which could arise from stray birefringence or variations in
the direction of an external magnetic field. Solutions to the
optical Bloch equations involving magnetic sublevels of the 3 lowest
states of the calcium ion show that projection errors can be kept
below 0.01 provided the relative Rabi frequencies of the unwanted
polarization components can be kept below 0.02-0.04.

A simulation including all effects outlined above, except
polarization errors, show the importance of being able to
efficiently switch on and off laser power as well as using highly
coherent or perhaps phase locked lasers in this qubit detection
scheme. Since we assume that the lasers involved in the scheme are
only partially coherent, a loss of transfer efficiency of almost
0.02 is found for laser linewidths of 2 kHz over the time of the
experiment. Performing the shelving on a shorter timescale will
typically reduce the relevant linewidths but also require higher
Rabi frequencies and consequently better suppression of residual
laser light.

\appendix*
\section{Relevant ion data}
\label{sec:append} Table \ref{tab:iondata} shows spectroscopic data
for some selected ions possessing the level structure and
transitions shown in Fig. \ref{fig:femniv}. Listed are wavelengths,
$\lambda_i$, associated with the fields $\Omega_i$, $i=\lbrace
A,B,C,D \rbrace$, decay rates of the short lived states $\vert
2\rangle$ and $\vert 4\rangle$ into the stable and metastable states
as well as decay rates of the metastable states into the stable
states and the Doppler temperature, $T_D$. The data is reproduced
from Ref. \cite{James98}.

\begin{table}[h!b!p!]
\caption{Data of relevant ions}
\begin{tabular}{lrrrr}

\hline\hline
\\[1mm]
\vspace{3pt}
Property & $^{40}$Ca$^+$ & $^{88}$Sr$^+$ & $^{138}$Ba$^+$ & $^{202}$Hg$^+$ \\
\hline
$\lambda_A$ [nm] & 397 & 422 & 493 & 194 \\
$\lambda_B$ [nm] & 866 & 1092 & 650 & 10670 \\
$\lambda_C$ [nm] & 850 & 1004 & 585 & 991 \\
\vspace{3pt}
$\lambda_D$ [nm] & 854 & 1033 & 614 & 398 \\
$\Gamma_{21} /2\pi$ [MHz] & 21 & 20 & 14 & 69 \\
$\Gamma_{23} /2\pi$ [MHz] & 1.7 & 1.5 & 5.3 & 0.05 \\
$\Gamma_{41} /2\pi$ [MHz] & 22 & 23 & 19 & 168 \\
$\Gamma_{43} /2\pi$ [MHz] & 0.18 & 0.18 & 0.76 & 0.48 \\
$\Gamma_{45} /2\pi$ [MHz] & 1.6 & 1.4 & 5.9 & 40 \\
\vspace{3pt}
$\Gamma_{31} /2\pi$ [Hz] & 0.15 & 0.40 & 0.009 & 1.62 \\
$\Gamma_{51} /2\pi$ [Hz] & 0.15 & 0.46 & 0.003 & 7.96 \\
\vspace{3pt}
$T_D$ [mK] & 0.50 & 0.49 & 0.35 & 1.7 \\
\hline
\end{tabular}
\label{tab:iondata}
\end{table}

\begin{acknowledgments}
This work has been supported by the Danish National Research Council
and by the Carlsberg Foundation.
\end{acknowledgments}

\bibliographystyle{apsrev}
\bibliography{stirap}

\end{document}